\documentclass[conference, a4paper]{IEEEtran}
\IEEEoverridecommandlockouts

\usepackage[font=footnotesize,caption=false]{subfig} 

\usepackage{cite}
\usepackage{amsmath,amssymb,amsfonts}
\usepackage{algorithmic}
\usepackage{graphicx}
\usepackage{textcomp}
\usepackage{xcolor}
\usepackage{booktabs}
\usepackage{glossaries}
\usepackage{subfig}
\usepackage{algorithm}
\usepackage{algorithmic}
\usepackage{bbm}
\usepackage{bm}
\usepackage{multirow}
\usepackage{tabularx}

\usepackage{tikz}
\usetikzlibrary{shapes.arrows}
\usetikzlibrary{plotmarks}
\usetikzlibrary{patterns}
\usetikzlibrary{arrows.meta}
\usetikzlibrary{positioning}
\usetikzlibrary{spy}

\usepackage{pgfplots}
\usepgfplotslibrary{colorbrewer}
\usepgfplotslibrary{fillbetween}
\newcommand{\set}{\mathcal}

\DeclareMathOperator*{\argmax}{argmax}

\newtheorem{remark}{Remark}

\def\BibTeX{{\rm B\kern-.05em{\sc i\kern-.025em b}\kern-.08em
    T\kern-.1667em\lower.7ex\hbox{E}\kern-.125emX}}
\newacronym{fec}{FEC}{forward error correction}
\newacronym{3gpp}{3GPP}{3rd Generation Partnership Project}    
\newacronym{iot}{IoT}{Internet of Things}
\newacronym{ntn}{NTN}{Non-Terrestrial Network}
\newacronym{leo}{LEO}{Low Earth Orbit}
\newacronym{geo}{GEO}{Geosynchronous Earth Orbit}
\newacronym{isl}{ISL}{Inter-Satellite Link}
\newacronym{gsl}{GSL}{Ground-to-Satellite Link}
\newacronym{qos}{QoS}{Quality of Service}
\newacronym{ofdma}{OFDMA}{Orthogonal Frequency-Division Multiple Access}
\newacronym{b5g}{B5G}{5G and Beyond}
\newacronym{mimo}{MIMO}{Multiple-Input Multiple-Output}
\newacronym{embb}{eMBB}{Enhanced Mobile Broadband}
\newacronym{urllc}{URLLC}{ultra-reliable and low-latency communications}
\newacronym{mmtc}{mMTC}{massive machine-type communications}
\newacronym{ue}{UE}{User Equipment}
\newacronym{nr}{NR}{New Radio}
\newacronym{gap}{GAP}{Generalized Assignment Problem}
\newacronym{mgap}{MGAP}{Multi-Level Generalized Assignment Problem}
\newacronym{csi}{CSI}{channel state information}    
\newacronym{ran}{RAN}{radio access network}
\newacronym{5g}{5G}{the 5th generation of mobile networks}
\newacronym{uav}{UAV}{unmanned aerial vehicle}
\newacronym{ap}{AP}{access point}
\newacronym{sic}{SIC}{successive interference cancellation}
\newacronym{mmimo}{mMIMO}{massive \gls{mimo}}
\newacronym{snr}{SNR}{signal to noise ratio}
\newacronym{sinr}{SINR}{signal to interference plus noise ratio}
\newacronym{fifo}{FIFO}{first-in first-out}
\newacronym{mab}{MAB}{multi-armed bandit}
\newacronym{rl}{RL}{reinforcement learning}
\newacronym{noma}{NOMA}{non-orthogonal multiple access}
\newacronym{rv}{RV}{random variable}
\newacronym{drl}{DRL}{deep \gls{rl}}
\newacronym{irsa}{IRSA}{irregular repetition slotted ALOHA}
\newacronym{oma}{OMA}{orthogonal multiple access}
\newacronym{fdma}{FDMA}{frequency division multiple access}
\newacronym{mdp}{MDP}{Markov decision process}
\newacronym{bs}{BS}{base station}
\newacronym{}{}{}

\begin{document}

\title{Heterogeneous radio access with multiple\\ latency targets }

\author{\IEEEauthorblockN{Israel Leyva-Mayorga\IEEEauthorrefmark{1}, Jose Manuel Gimenez-Guzman\IEEEauthorrefmark{2},  Lorenzo Valentini\IEEEauthorrefmark{1}\IEEEauthorrefmark{3}, and Petar Popovski\IEEEauthorrefmark{1}}
\IEEEauthorblockA{\IEEEauthorrefmark{1}Department of Electronic Systems, Aalborg University, Denmark (\{ilm,petarp\}@es.aau.dk)}
\IEEEauthorblockA{\IEEEauthorrefmark{2}Department of Communications, Universitat Polit\`ecnica de Val\`encia, Spain (jmgimenez@upv.es)}
\IEEEauthorblockA{\IEEEauthorrefmark{3}CNIT/WiLab, DEI, University of Bologna, Italy (lorenzo.valentini13@unibo.it)}
\thanks{
This work was partially supported by the Villum Investigator Grant \mbox{``WATER''} from the Velux Foundation, Denmark.}
}
\maketitle

\begin{abstract}
Since the advent of \gls{urllc}, the requirements of low-latency applications tend to be completely characterized by a single pre-defined latency-reliability target. That is, operation is optimal whenever the pre-defined latency threshold is met but the system is assumed to be in error when the latency threshold is violated. This vision is severely limited and does not capture the real requirements of most applications, where multiple latency thresholds can be defined, together with incentives or rewards associated with meeting each of them. Such formulation is a generalization of the single-threshold case popularized by \gls{urllc} and, in the asymptotic case, approximates to defining a cost for each point in the support of the latency distribution. In this paper, we explore the implications of defining multiple latency targets on the design of access protocols and on the optimization of repetition-based access strategies in orthogonal and non-orthogonal multiple access scenarios with users that present heterogeneous traffic characteristics and requirements. We observe that the access strategies of the users can be effectively adapted to the requirements of the application by carefully defining the latency targets and the associated rewards.
\end{abstract}
\glsresetall
\section{Introduction}
The coexistence of diverse services with heterogeneous requirements is a fundamental feature of 5G, which aims to serve as a connectivity platform capable of supporting users that require low latency, as well as users that require high data rates. To achieve low latency in the \gls{ran}, the 5G standards now include several signaling and resource allocation mechanisms~\cite{TS38.300,TS38.213, TS38.321}. However, achieving low latency in the uplink is still a great challenge in 5G, since most of its grant-based access mechanisms are inefficient when there is a high uncertainty on the activation of the users. On the one hand, long-term allocation of resources to a single user that might be inactive a large fraction of the time leads either to resource wastage.  On the other hand, users might experience a long access latency if the frequency of the allocated resources is low and it has to wait for a long time until the next allocated resource to transmit~\cite{Cavallero23}. In addition, sending a scheduling request every time new data is available for transmission results in a long access delay and increased communication overhead~\cite{Sachs2018}. Hence, it is in the uplink that the coexistence of grant-based and grant-free access mechanisms in time, frequency, and or code domains might provide the flexibility needed to support heterogeneous services with maximum resource efficiency. This is still not possible in 5G NR because the resources must be allocated to a single user following a grant-based access approach and non-orthogonal resource sharing between users is not possible. 

We argue that achieving efficient grant-free communication requires flexible and efficient \emph{\gls{ran} slicing}, defined as the sharing of the wireless resources among diverse services to meet their throughput, latency, and reliability requirements. In previous work, we investigated the coexistence of heterogeneous users with grant-based and grant-free mechanisms~\cite{Chiariotti2022} but our analyses did not consider practical feedback mechanisms and, instead, assumed perfect and immediate feedback.

\begin{figure}[t]
    \centering
\includegraphics{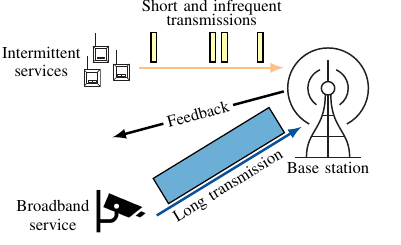}\vspace{-0.5em}
    \caption{Uplink scenario with one user running a broadband service and multiple \gls{iot} users running intermittent services, which generate small amounts of data sporadically. The \gls{bs} provides feedback, which can be used to design the access policies, including the initial data transmission and retransmissions.}
    \label{fig:diagram}
    \vspace{-1em}
\end{figure}

In this paper, we investigate the uplink scenario shown in Fig.~\ref{fig:diagram}, where a \gls{bs} slices the \gls{ran} resources among one user generating broadband data and one or more users with intermittent and random data transmissions, hence referred to as \emph{intermittent users}. In contrast to the rigid definition of low latency, the requirements of the intermittent users are represented by multiple latency targets (i.e., thresholds), where the cost of violating subsequent thresholds increases. This model is applicable to systems with an application or actuator whose performance degrades as the packet latency increases. To promote meeting the latency targets, the application, controller, and/or the network infrastructure define a set of rewards, which serve as incentives for the users with the aim of finding appropriate access strategies to maximize the expected total reward. Thus, the objective is to jointly design the data transmission strategies and immediate rewards for the given latency targets and frame scheduling (i.e., placement of the feedback, initial transmission, and retransmissions phases) to maximize the expected rewards for the whole transmission process. In particular, we focus on finding optimal transmission policies for repetition-based access mechanisms, where the action of the users with intermittent activation is the placing of a given number of repetitions within a frame. Optimizing the number of repetitions has been a topic of high interest but model-based optimizations have proven to be intractable in numerous scenarios. For this reason, model-free centralized optimization \gls{drl} has been investigated~\cite{Ayoub21}. An interesting direction is considering not only the maximization of the number of successful accesses but also the energy efficiency of the access mechanisms~\cite{Jia19}.

\section{System model}
We consider an uplink scenario where a broadband user and an intermittent user are allocated resources for communication with a \gls{bs}. These users are indexed by $m$, where $m=1$ is the broadband user and $m=2$ is the intermittent user. The wireless resources are located in a frequency band with total bandwidth $B$~Hz. The available bandwidth $B$ is partitioned into three sub-bands, each with bandwidth: 1) $B_1$ reserved for the broadband user; 2) $B_2$ reserved for the intermittent user; and 3) $B_3$ shared by both users, such that $B=B_1+B_2+B_3$. To indicate the allocation of users to sub-bands, let $\alpha_{m,i}\in\{0,1\}$ be the indicator function, which takes the value of $1$ if user $m\in\{1,2\}$ is allocated to sub-band $i\in\{1,2,3\}$. Building on this, two slicing mechanisms are defined:
\begin{enumerate}
\item \emph{\Gls{fdma}:} Following an \gls{oma} approach, the users are allocated resources in non-overlapping parts of the available bandwidth, with $B_1$ containing slots reserved for the broadband user and $B_2$ containing slots reserved for the intermittent user. This is achieved by setting $\alpha_{1,1}=\alpha_{2,2}=1$, $B=B_1+B_2$, and $B_3=0$, such that there is no signal overlap between the users.
\item \glsreset{noma}\emph{\Gls{noma}:} Both users are allowed to use the whole bandwidth by setting $\alpha_{1,3}=\alpha_{2,3}=1$, $B_3=B$, and $B_1+B_2=0$, which leads to signal overlap when both transmit in the same time slot.
\end{enumerate}

\begin{figure}
\centering
\includegraphics{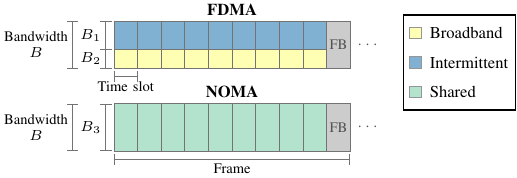}\vspace{-0.2em}
\caption{\Gls{ran} slicing between a broadband and an intermittent user in a time-slotted frame with feedback (FB) in the last slot. In \gls{fdma}, the bandwidth $B$ is split orthogonally among the users: $B_1$ for the broadband and $B_2$ for the intermittent user. In \gls{noma}, both users can access the whole bandwidth.}
\label{fig:multiple_access}
\vspace{-1em}
\end{figure}

Communication takes place over a slotted channel with slot duration $T_s$ and where communication is segmented into frames with $F$ consecutive time slots. The first $F-1$ slots are used  for uplink transmission and the last slot in a frame is reserved for downlink communication, where the \gls{bs} provides feedback (i.e., ACKs) to the users. The frame structure and the slicing in both \gls{fdma} and \gls{noma} are illustrated in Fig.~\ref{fig:multiple_access}.

The broadband user segments the data to transmit into packets and implements a \gls{fec} mechanism with an ideal rateless packet-level coding, where blocks of $K$ source packets are encoded into blocks of linearly independent packets. Therefore, a block is decoded by the \gls{bs} when $K$ packets belonging to the same block are received.

The intermittent user generates a new packet of length $L$ with probability $p_a$ at each time slot. We consider that the maximum length of the transmission queue of the intermittent user is $1$ and, hence, the user discards any new arrivals that occur when a previously generated packet is being transmitted. These packets are transmitted to the \gls{bs} in the following frame according to a repetition-based grant-free access protocol using either sub-band $2$ or $3$. The number of repetitions to be transmitted by the intermittent user at frame $f\in\{0,1,\dotsc\}$, oftentimes called the \emph{repetition degree}, is denoted as $a$. The application at the end receiver defines an ordered list of $\tau$ latency targets $\set{L}=\left(\ell^{(1)},\ell^{(2)}, \dotsc,\ell^{(\tau)}\right)$, where meeting the latency target $\ell^{(j)}$ results in a better performance than meeting any of the $\tau-j$ higher targets. 

The users in the system communicate over a block fading channel with capture, where the transmissions are affected by path loss and fast fading with coherence time equal to one time slot. Each user selects its own nominal transmission rate $r_m$ that remains fixed throughout the operation of the system. 

 Let $h_{m,t}$ and $x_{m,t}$ be the channel coefficient and the transmitted signal at time slot $t$ by user $m$. Consequently, the channel model for uplink communications at the $i$-th sub-band, for $i\in\{1,2,3\}$, at the $t$-th time slot is
\begin{equation}
\vspace{-0.4em}
y_{i,t} =  h_{1,t}\,x_{1,t}\,\alpha_{1,i} + h_{2,t}\,x_{2,t}\,\alpha_{2,i} + w_{i,t},
\vspace{-0.4em}
\end{equation}
where $w_{i,t}$ is the circularly-symmetric Gaussian noise. Let $\kappa$ be the Boltzmann constant, $T_n$ be the system noise temperature, and $N_\text{f}$ be the noise figure in dB. Consequently,  the noise power is $\sigma_i^2=B_i\kappa T_n\, 10^{N_\text{f}/10}$. Next, let $P_{m,t}$ be the transmission power of user $m$ at time slot $t$, where $P_{m,t}\in\left(0, P_\text{max}\right]$ if user $m$ is active at time slot $t$ and $P_{m,t}=0$ otherwise. The channel coefficient $h_{m}$ is a random variable affected by path loss and fast fading. Thus, the expected value of the channel coefficient for a user $m$ is a function of the carrier frequency $f_c$, the transmitter and receiver antenna gains, $G_m$ and $G_r$, respectively, the distance from the user to the \gls{bs} $d_m$, and the path loss exponent $\eta$, namely,
\begin{equation}
\vspace{-0.2em}
    \mathbb{E}\left[|h_{m}|^2\right] =\frac{G_m\,G_r\,c^2}{\displaystyle(4\pi f_c)^2\,d_m^\eta},
\vspace{-0.2em}
\end{equation}
where $c$ is the speed of light. Building on this, the \gls{sinr} for the signal transmitted by user $m$ in sub-band $i$ at time slot $t$, which can be affected by interference from user $n\neq m$, is defined as
\begin{equation}
    \gamma_{m,i,t} = \frac{|h_{m,t}|^2\,P_{m,t}\alpha_{m,i}}{\displaystyle \sigma_i^2+ |h_{n,t}|^2\,P_{n,t}\alpha_{n,i}}.
    \label{eq:sinr}
\end{equation}
We assume that interference can be treated as Gaussian noise and define a threshold for decoding the signal of user $m$ in sub-band $i$ at time slot $t$ as the function of $r_{m}$ given as
\begin{equation}
    \gamma_{m,i}^\text{min}\left(r_{m}\right) = 2^{\displaystyle r_{m}/B_i}-1,
\end{equation}
such that the packet from user $m$ is successfully decoded at time slot $t$ if $\gamma_{m,i,t}\geq \gamma_{m,i}^\text{min}\left(r_{m}\right)$. The erasure probability for user $m$, denoted as $\epsilon_{m,i}$, is calculated as the probability of a transmission error in the absence of interference from the other user $n\neq m$, namely,
\begin{equation}
    \epsilon_{m,i} = \Pr\left(\gamma_{m,i,t}<\gamma_{m,i}^\text{min}\left(r_{m}\right): \alpha_{m,i}=1\wedge P_{n,t}\,\alpha_{n,i}=0\right).
\end{equation}
However, in the general case, a packet transmitted by user $m$ in sub-band $i$ at time $t$ is successfully decoded with probability
\begin{equation}
    p_{m,i,t} = \Pr\left(\gamma_{m,i,t}\geq\gamma_{m,i}^\text{min}\left(r_{m}\right)\right).
    \label{eq:p_succ}
\end{equation}
The broadband user $m=1$ selects its transmission power $P_{1,t}$ based on the desired transmission rate $r_{1}$ and the target erasure probability $\epsilon^*_1$, which can be calculated from~\eqref{eq:p_succ} by removing the interference from~\eqref{eq:sinr}. Hence, by considering a Rayleigh fading channel and that the broadband user has statistical knowledge of the channel coefficient $h_{1}$, it selects the transmission power as
\begin{equation}
    P_{1,t} =
        \min\left(\frac{\left(2^{\displaystyle r_{1}/B_i}-1\right)\,\sigma_i^2}{ \displaystyle\mathbb{E}\left[|h_{1}|^2\right]\log(\epsilon^*_1-1)}, P^\text{max}\right).
    \end{equation}
On the other hand, we assume that the intermittent user has no statistical knowledge of $h_{2}$ and selects the maximum transmission power $P_{2,t}=p^\text{max}$ when active. 

Furthermore, we model the energy consumption for the transmission of a packet by user $m$ at time $t$ as
\begin{equation}
    E_{m,t}=T_s\, P_{m,t}.
    \label{eq:econs}
\end{equation}

We consider a realistic \gls{sic} decoder, in which the \gls{bs} attempts to decode the signals received in a time slot immediately at the end of it. The success or failure of the decoding process is based on the \gls{sinr} defined in~\eqref{eq:sinr}. Given that the users' signals are subject to independent Rayleigh fading, it is not possible for the \gls{bs} to know which of the two signals will have a higher \gls{sinr} before these are decoded. Therefore, the decoder attempts to decode both users at each time slot, for example, by attempting to decode the signal from user $1$ first, followed by that of user $2$. We assume that, after decoding the signal from a user $m$, obtaining $h_{m,t}x_{m,t}\alpha_{m,i}$ is possible. Hence, the interference of the decoded signal is completely removed from~\eqref{eq:sinr} and, then, the \gls{bs} attempts to decode the other user.  

 At the end of the frame, the \gls{bs} sends a feedback message, which indicates the \emph{state} of the system and is used by the users to select their following \emph{action}. For the broadband user, the feedback from the \gls{bs} indicates whether the current block has been decoded successfully or not. Then, the broadband user proceeds to transmit the next block only when the reception of the current block is acknowledged through feedback. Note that, by following this transmission strategy, the broadband user is guaranteed to achieve reliability $1$ in the cases where $p_{1,i,t}>0$. Thus, the broadband user selects the rate $r_1$ to be the minimum between a maximum data rate $r_{1}^\text{max}$ and the maximum achievable rate for the target erasure probability $\epsilon^*_1$
\begin{equation}
    r_1=\max\big\{r\in\left(0,r_1^\text{max}\right]: \epsilon_{1,i}(r)=\epsilon^*_1, P_{1,t}\leq P^\text{max}\big\}.
\end{equation}

Next, let $W(K)$ be the \gls{rv} of the number of frames needed to decode the block of $K$ source packets transmitted by the broadband user. Building on this, the throughput of the broadband user is
\begin{equation}
S= r_1\,K/\left(\mathbb{E}\left[W(K)\right]\,F\right)
\end{equation}
and its energy efficiency is
\begin{equation}
    \text{EE} = S/P_{1,t}.
\end{equation}

 On the other hand, we model the access of the intermittent user as a \gls{mdp} with state space $\set{S}$, action space $\set{A}$, and rewards $\set{R}$. The state of the intermittent user at a given time slot $t$ is defined by the tuple $s_t=(\ell, k,\delta)$, where $\ell$ is the latency of the current packet in the transmission queue, $k$ is the total number of repetitions performed since the generation of the current packet in the transmission queue, and $\delta$ indicates whether the packet has been decoded by the \gls{bs}, with $\delta=1$ indicating correct decoding and $\delta=0$ otherwise. 
 
 At the beginning of each frame $f\in\{0,1,\dotsc\}$, the intermittent user selects an action $a\in\set{A}=\{0,1,\dotsc, F-1\}$, which indicates the repetition degree to be used in the current frame. To minimize latency, the repetitions are placed one after the other in consecutive time slots, starting from the first one in the frame. If two or more repetitions of a packet are successfully decoded in the same frame, $\ell$ is set to the latency of the first decoded repetition. Otherwise, the latency $\ell$ increases at each time slot. However, the intermittent user can only observe the state of the system at the end of each frame, immediately after receiving the feedback and before choosing the action for the next frame. Once the successful decoding of a packet is indicated by the feedback or the maximum latency $\ell^{(\tau)}$ is exceeded, the packet is removed from the transmission queue of the intermittent user. Consequently, the state of the intermittent user transitions to state $(0,0,0)$ immediately after a successful transmission and also after the maximum latency is exceeded, which results in a failed transmission attempt.

Let $R(\ell)\in\set{R}$ be the immediate reward for a successful transmission from the intermittent user with latency $\ell$. The immediate reward is defined based on the ordered list of latency targets $\set{L}=\left(\ell^{(1)}, \ell^{(2)}, \dotsc, \ell^{(\tau)}\right)$, where $\ell^{(j)}>\ell^{(j-1)}$ for all $j\in\{2,3,\dotsc,\tau\}$. Specifically, the immediate reward $R(\ell)$ is the decreasing step function 
\begin{equation}
    R(\ell) = \max\left\{R\left(\ell^{(j)}\right)\in\set{R}:\ell\leq \ell^{(j)}\right\}. 
        \label{eq:imm_rew}
\end{equation}
Finally, the discounted reward for a given state $(\ell, k, \delta)$ based on the immediate reward~\eqref{eq:imm_rew} and the total number of repetitions for the current packet, namely
\begin{equation}
    R(\ell, k, \delta)=\begin{cases}
        \displaystyle R(\ell)-k,  & \text{if } \delta=1 \text{ or } \ell>\ell^{(\tau)} \\
        0, & \text{otherwise}.
    \end{cases}
    \label{eq:reward}
\end{equation}

If the success probability per transmission for the intermittent user $p_{2,i,t}$ and the rewards for each state $R(\ell, k, \delta)$ are known, the optimal transmission policy for the intermittent user $\bm{\pi}$ can be easily obtained by value iteration\cite{SuttonBarto}. Specifically, the transition probabilities from a given state $s_t\in\set{S}$, where $t$ is the first time slot of a frame, to any given state $s_{t+F}\in\set{S}$ given an action $a\in\set{A}$, denoted as $p\left(s_{t+F}\mid s_t, a\right)$ are fully characterized by $p_{2,i,t}$. Furthermore, the reward for a given state $s_{t+F}$ is fully characterized by $R(\ell, k, \delta)$ and the value of a state $s\in\set{S}$ can be updated iteratively by applying
\begin{equation}
    V(s)^{v+1} = \max_a \sum_{s'\in\set{S}} p\left(s'\mid s,a\right)\left(R(s')+\beta V(s')^v\right),
    \label{eq:value}
\end{equation}
where $v$ is the current iteration, $\beta$ is a discount factor that controls the importance of short-term rewards, and the state values are initialized to $V(s)^0=0$ for all $s\in\set{S}$. Convergence is achieved by applying~\eqref{eq:value} until 
\begin{equation}
    \max_{s\in\set{S}} \left|V(s)^{v+1}-V(s)^v\right|\leq \varepsilon,
\end{equation}
where $\varepsilon>0$ is the pre-defined tolerance. Then, the optimal policy $\bm{\pi}^*=\{\pi^*(s)\}$, mapping a deterministic action $a$ for all $s\in\set{S}$, is derived from the final values for the states $V^*(s)$ as
\begin{equation}
    \pi^*(s)=\argmax_a \sum_{s'\in\set{S}} p\left(s'\mid s,a\right)\left(R(s')+\beta V^*(s')\right).
\end{equation}
\begin{remark} Since value iteration typically converges after a few iterations and does not incur in a high computational load, the proposed solution can be implemented by estimating $p_{m,i,t}$ and calculating the optimal values either 1) at the end device by and receiving the latency targets $\set{L}$ and the associated immediate rewards $\set{R}$ from the \gls{bs} or 2) at the \gls{bs} itself.
\end{remark}
\begin{remark} Given that the transmission power for the intermittent user is known to be always $P_\text{max}$, it is possible to define $R(\ell)$ to serve as a weight that balances the benefits of transmitting a packet with latency $\ell$ and the cost of transmission (e.g., in terms of energy consumption). Specifically, setting relatively low values for $R(\ell)$ would lead to an access policy that aims to minimize energy consumption by transmitting a low number of repetitions. On the other hand, setting relatively high values for $R(\ell)$ would lead to an access policy that aims to maximize the reliability for the latency targets by transmitting a high number of repetitions. 
\end{remark}

\section{Results}
In this section, we evaluate the performance of the proposed policy optimization approach through value iteration on \gls{fdma} and \gls{noma} frames. The broadband user is located at $50$~m from the \gls{bs}, whereas the intermittent user is located at different distances $d$ from the \gls{bs}. In the baseline scenario, the list of latency targets is $\set{L}=\left(20,40\right)$ and the list of rewards is $\set{R}=\left(10,3\right)$. The results are obtained by simulation, with each simulation containing at least $100000$ frames. The parameter settings are listed Table~\ref{tab:parameters}.

\begin{table}[t]
    \centering
    \renewcommand{\arraystretch}{1.1}
        \caption{Parameter settings for performance evaluation.}\vspace{-1em}
    \begin{tabularx}{\columnwidth}{@{}Xll@{}}
    \toprule
    Parameter & Symbol & Setting  \\
    \midrule
   Erasure probability of the broadband user & $\epsilon^*_1$ & $0.1$\\
    Maximum data rate of the broadband user & $r_1^\text{max}$ & $5$~Mbps\\
Maximum transmission power & $P_\text{max}$  & $200$~mW\\
Antenna gains & $G_tG_r$ & $10$\\
Time slot duration & $T_s$ & $1$~ms\\
Packet size for intermittent users & $L$ & $128$~B\\
Carrier frequency & $f_c$ & $2$~GHz\\
System bandwidth & $B$ & $1$~MHz\\
Noise temperature & $T_n$ & $190$~K\\
Noise figure & $N_f$ & $5$~dB\\
Frame length & $F$ & $10$\\
Source block length for the broadband user & $K$ & $32$\\
Distance between the users and the \gls{bs} & $d$ & $\left\{100,200,400\right\}$~m\\
Path loss exponent & $\eta$ & $2.6$\\
\bottomrule
    \end{tabularx}
    \label{tab:parameters}
    \vspace{-1em}
\end{table}

\begin{figure}[t]
\subfloat[]{\includegraphics{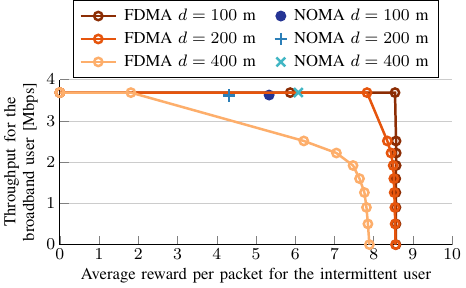}\label{fig:th_trade-off}}\vspace{-0.5em}\\
\subfloat[]{\includegraphics{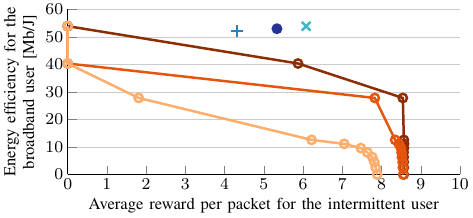}\label{fig:ee_trade-off}}\vspace{-0.2em}\\
\caption{Trade-off between the average reward for the intermittent user and (a) the throughput and (b) the energy efficiency of the broadband user given $\set{L}=(20,40)$ and $\set{R}=(10,3)$.}
\label{fig:trade-off}
\vspace{-1em}
\end{figure}

Fig.~\ref{fig:trade-off} shows the achievable trade-offs between the average reward per packet for the intermittent user and two metrics for the broadband user: (a) throughput and (b) energy efficiency for $d=\{100,200,400\}$~m. We begin by investigating the behavior of \gls{fdma}, for which the different points in Fig.~\ref{fig:trade-off} are obtained by setting $B_1\in\{0,B/10,\dotsc,B\}$. Naturally, the maximum throughput for the broadband user is obtained by \gls{fdma} with $B_1=B$ but this also results in $0$ reward for the intermittent user. It is also evident that the performance with \gls{fdma} increases as the distance between the intermittent user and the \gls{bs} decreases. Nevertheless, a negligible throughput degradation is observed by slightly reducing $B_1$ to accommodate the transmissions of the intermittent user. However, reducing $B_1$ requires an increase in transmission power of the broadband user to maintain the nominal data rate of $r_1^\text{max}=5$~Mbps. Hence, if $B_1$ is considerably reduced, $P_\text{max}$ becomes insufficient to maintain $r_1^\text{max}$ for the target erasure probability $\epsilon^*_1$ and the broadband user must reduce the nominal data rate, which leads to the drop in throughput observed in the right part of Fig.~\ref{fig:th_trade-off}. The need for the broadband user adapting the transmission power and the nominal rate is clearly observed in Fig.~\ref{fig:ee_trade-off}, where the energy efficiency of the broadband user must be decreased in exchange for increasing the reward of the intermittent user. 

Next, we focus on the behavior of \gls{noma}, which exhibits two interesting aspects. First, both the throughput (see Fig.~\ref{fig:th_trade-off}) and energy efficiency (see Fig.~\ref{fig:ee_trade-off}) achieved by the broadband user are similar to the maximum values obtained with \gls{fdma}, while the reward for the intermittent user is modest, neither reaching the maximum obtained with \gls{fdma} nor being close to $0$. In particular, the combination of a high energy efficiency and modest reward places the results with \gls{noma} outside of the feasible region of \gls{fdma}. Second, the reward achieved by the intermittent user located at $d=200$~m is the lowest. This behavior is typical in \gls{noma}, since small differences between the \glspl{snr} of the users lead to low capture probabilities. Thus, it can be concluded from Fig.~\ref{fig:trade-off} that pairing broadband users close to the \gls{bs} with an intermittent user at a larger distance (e.g., $d=400$~m) results in the best performance.

\begin{figure}[t]
\subfloat[]{\includegraphics{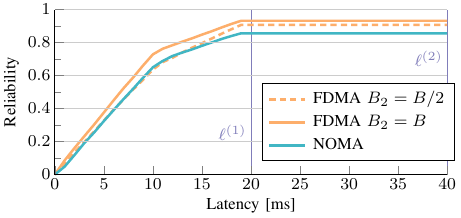}\label{fig:lat_cdf_baseline}}\vspace{-0.5em}\\
\subfloat[]{\includegraphics{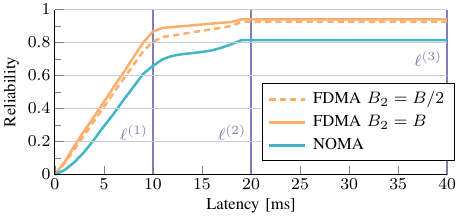}\label{fig:lat_cdf_lowlatency}}\vspace{-0.2em}
\caption{Latency-reliability function for the packets generated by an intermittent user at $d=400$~m from the \gls{bs} under \gls{fdma} and \gls{noma} given (a) $\set{L}=\left(20,40\right)$ and $\set{R}=\left(10,3\right)$ and (b) $\set{L}=\left(10,20,40\right)$ and $\set{R}=\left(20,10,3\right)$.}
\label{fig:lat_cdf}
\vspace{-1em}
\end{figure}

To conclude, we analyze the impact of different latency targets $\set{L}$ and rewards $\set{R}$ on the latency-reliability function (i.e., latency CDF). Fig.~\ref{fig:lat_cdf_baseline} shows the latency-reliability for the baseline scenario and Fig.~\ref{fig:lat_cdf_lowlatency} shows the one for the case with an additional latency target of $10$~ms and an additional reward of $R(10)=20$. Three curves are shown for each scenario: 1) \gls{fdma} with $B_1=B_2=B/2$, 2) the upper bound in performance with \gls{fdma} and $B_2=B$, and \gls{noma}. Clearly, the optimal policies for the three cases aim to achieve a reliability higher than $0.8$ for a latency of $20$~ms, which results in a high reward. However, since the reward for achieving a latency between $20$ and $40$ is comparatively low, the increase in the latency-reliability functions in this latency range is negligible. On the other hand, providing a high reward for transmitting with a latency lower than $10$~ms results in a marked increase in the latency-reliability function shown in Fig.~\ref{fig:lat_cdf_lowlatency}, which showcases the ability to shape the latency-reliability function of the intermittent users by correctly defining the latency targets and their associated rewards.

\section{Conclusion}
In this paper, we considered the problem of slicing the \gls{ran} resources in the frequency domain among a broadband and an intermittent user for uplink transmission. Furthermore, we showcased how defining multiple latency targets and their associated rewards can be used to shape the latency-reliability function of the intermittent user. In addition, we observed that the optimal policy can be obtained by applying value iteration if the success probability of the transmissions is known or can be accurately estimated. Nevertheless, our analyses are limited to one intermittent and one broadband user. Accordingly, future work includes the extension of the proposed mechanisms and analyses to the case with multiple intermittent users. In particular, the applicability of value iteration to multiple intermittent users and the need for model-free reinforcement learning techniques must be investigated.
\bibliographystyle{IEEEtran}
\bibliography{bib}
\end{document}